\documentclass[aps,prl,showpacs,twocolumn,superscriptaddress]{revtex4}
\usepackage{mathrsfs}
\usepackage{epsfig}
\usepackage{graphicx}
\usepackage{amsfonts}
\usepackage[figuresright]{rotating}
\usepackage{amssymb}
\usepackage{amsmath}
\usepackage{dcolumn}
\usepackage{bm}
\usepackage{color}

\def\be{\begin{equation}} \def\ee{\end{equation}}
\def\bea{\begin{eqnarray}} \def\eea{\end{eqnarray}}

\newcommand{\IFR} {Facultad de Ciencias Exactas y Naturales, Universidad Nacional de Cuyo and CONICET, Mendoza 5500, Argentina}

\newcommand{\WQCASQC} {Wilczek Quantum Center and Key Laboratory of Artificial Structures and Quantum Control, School of Physics and Astronomy, Shanghai Jiao Tong University, Shanghai 200240, China}

\begin{document}
\title{Dissipative Majorana quantum wires}

\author{Yizhen Huang}
\affiliation{\WQCASQC}

\author{Alejandro M. Lobos}
\affiliation{\IFR}

\author{Zi Cai}
\email{zcai@sjtu.edu.cn}
\affiliation{\WQCASQC}

\begin{abstract}
In this paper, we formulate and quantitatively examine the effect of dissipation on topological systems. We use a specific model of Kitaev quantum wire with an onsite Ohmic dissipation, and perform a numerically exact quantum Monte Carlo simulation to investigate this interacting open quantum system with a strong system-bath (SB) coupling beyond the scope of Born-Markovian approximation. We concentrate on the effect of dissipation on the topological features of the system (e.g. the Majorana edge mode) at zero temperature, and find that even though the topological phase is robust against weak SB couplings as it is supposed to be,  it  will eventually be destroyed by sufficiently strong dissipations via either a  continuous quantum phase transition or a crossover depending on the symmetry of the system. The dissipation-driven quantum criticality is also discussed. In addition, using the framework of Abelian bosonization, we provide an analytical description of the interplay between pairing, dissipation and interaction in our model.

\end{abstract}

\maketitle

{\it Introduction:} Topological quantum phases of matter are among the most  notable phenomena in condensed matter physics\cite{Thouless1998}. Instead of being classified by symmetries and their spontaneous breaking,  topological phases of matters are identified by nonlocal topological orders that are immune to local perturbations\cite{Wen2004}. The intrinsic stability of the topological features in the underlying systems makes them a promising platform for quantum computation and information processing\cite{Nayak2008}. One of the major obstacles for the realization of a practical quantum computer is that quantum systems are inevitably coupled to their surroundings, which gives rise to dissipation and decoherence that is detrimental to the quantum coherence\cite{Schlosshauer2007}.  Since coupling to the environment tends to drive a quantum system to be classical, while topological phases are quantum in nature,  it is natural to expect that sufficiently large bath-induced dissipation and decoherence will eventually destroy the topological phases is spite of their robustness against small perturbations.  The question is: How large? And whether the system experiences a crossover or a phase transition during this process?  Understanding a topological system immersed in an environment is not only of fundamental interest of topology physics itself,  but also of immense practical significance in quantum simulation and information processing; and hence deserves quantitative studies rather than qualitative arguments.

Formulating and quantitatively examining the problem poses multiple challenges: the first one is properly dealing with the environment. Most researchers follow the quantum optics strategy\cite{Breuer2002} of tracing out the bath degrees of freedom and deriving a Born$-$Markov master equation for the open quantum systems with a weak system-bath (SB) coupled to a Markovian environment\cite{Diehl2011b,Bardyn2012,Budich2015,Linzner2016}. This strategy doesn't apply to our case where the SB couplings are intrinsically strong\cite{Hanggi2006,Hanggi2008} and the environment in solid-state devices generally is non-Markovian ({\it e.g.} phonon-coupling). Secondly, topological phases are usually formulated in terms of  pure (ground) state, while open quantum systems are in general described by mixed states. Generalization of topological phases to mixed states is a non-trivial problem and a variety of theoretical efforts have been employed\cite{Uhlmann1986,Viyuela2014a,Viyuela2014b,Huang2014,Budich2015b,Bardyn2018}, while most of them focus on  non-interacting systems and only a few are known for the interacting cases\cite{Grusdt2017,Trebst2007}. This immediately leads to the third challenge: an open system with strong SB coupling is usually a genuine interacting system, even though the system Hamiltonian itself is noninteracting ({\it e.g.} a topological insulator), since the bath may induce effective interactions between the system particles  ({\it e.g.} the attractive interactions in superconductors). Understanding the topological phases in interacting quantum systems is challenging, what further complicates the problem is that the environment-mediated interactions are usually time-delayed if the bath is non-Markovian, while few numerical methods currently used in strongly correlated physics can be applied to the systems with retarded interactions. In summary, we are in  face of a topological quantum many-body system strongly coupled to a non-Markovian bath, which alters the interactions within the system to be time-delayed.

In this paper, we investigate the fate of a topological phase in the presence of a strong dissipation by performing a numerically exact Quantum Monte Carlo(QMC) simulation as well as an analysis using Abelian bosonization method. We model our system Hamiltonian as a Kitaev wire composed of spinless fermions\cite{Kitaev2001}, a prototypical example to illustrate nontrivial topology and edge state in one-dimensional(1D) lattice geometry, while the environment is modeled by sets of harmonic oscillators with Ohmic spectrum following Caldeira-Leggett's seminal work\cite{Caldeira1981, Caldeira1983,Caldeira1983b,Leggett1987}. The fermions in the Kitaev wire couple to the bath via their density operators.   The key outcome of this paper is that, except for a special point, the system experiences a continuous quantum phase transition (QPT) from a topological nontrivial phase to a strongly dissipative phase with increasing dissipation. The fate of Majorana fermions in the presence of dissipation has  also been investigated.

{\it Model and method --}
 The Hamiltonian of a dissipative  system contains three parts and is expressed as $H_{tot}=H_s+H_b+H_{sb}$. $H_s$ is the system Hamiltonian chosen as a Kitaev wire and is given as follows:
\begin{equation}
H_s=\sum_{\langle ij\rangle}\{-J(c_i^\dag c_j+c_j^\dagger c_i)-\Delta(c_i^\dag c^\dag_j+c_j c_i) \}-\mu \sum_i n_i, \label{eq:Kitaev}
\end{equation}
where $c_i$ ($c_i^\dag$) are the annihilation (creation) operators of spinless fermions at site $i$,  $J$ ($\Delta$) denotes the hopping (pairing) amplitude between nearest neighboring sites and $\mu$ the chemical potential. Without loss of generality, we choose $\Delta=J$ in the following. On each site $i$, a fermion additionally couples to a local bath (modeled by a set of harmonic oscillators) via its density operator $n_i$. The Hamiltonians describing each local bath and system-bath coupling read as follows:
\begin{eqnarray}
H_b&=&\sum_{i,k}  \frac{P_{ik}^2}{2m_k}+\frac 12 m_k\omega_k^2 X_{ik}^2,\\
H_{sb}&=&\sum_{i,k} [\frac{\lambda_k}{\sqrt{2m_k\omega_k}} (n_i - \frac{1}{2}) X_{ik}], \label{eq:sbHam}
\end{eqnarray}
where $X_{ik}$ ($P_{ik}$) denotes the coordinate(momentum) operator of the bath harmonic oscillator with modes $\omega_k$ on site $i$.  The baths around different system sites are independent of each other, but are characterized by the same Ohmic spectral function:
$J(\omega)=\pi\sum_k \frac{\lambda_k^2}{2m_k\omega_k} \delta(\omega-\omega_k)=\pi\alpha
\omega$ for $0<\omega< \omega_D$ and $J(\omega)=0$ otherwise.  $\omega_D$ is a hard frequency cutoff chosen as $\omega_D=10J$ and $\alpha$ is the dissipation strength.

Integrating out the bath degrees of freedom leads to a retarded  interaction term in imaginary time. The total system (system+bath) is assumed to be in thermal equilibrium at temperature $T=1/\beta$, thus the partition function of the total system takes the form $Z={\rm Tr} e^{-\beta H_{tot}} =Z_B \times {\rm Tr}_s \rho_s$ where  $Z_B$ is the partition function for the free bosons of the bath, $\rho_s$ is the reduced density matrix of the system and take the form given below
\begin{equation}
\rho_s= e^{-\beta H_s+\int_0^\beta d\tau \int_0^\beta d\tau' \sum_i
(n_i (\tau) - \frac{1}{2}) D(\tau-\tau') (n_i(\tau') - \frac{1}{2})}. \label{eq:density}
\end{equation}
The effect of dissipation is encapsulated in the onsite retarded interaction in Eq.(\ref{eq:density}) characterized by the site-independent kernel function of the Ohmic spectrum~\cite{Winter2009} $D(\tau)=\int_0^\infty d\omega \frac{J(\omega)}{\pi} \frac{\cosh(\frac{\omega\beta}2-\omega|\tau|)}{\sinh(\frac{\beta\omega}2)}$. In the limit of $T=0$ and  $\tau\gg \tau_c=2\pi/\omega_D$, $D(\tau) \sim  1/\tau^2$.  The reason for the choice of the factor $\frac 12$  in Eq.(\ref{eq:sbHam})  is that we wish the bath effect to be purely dynamical, such that the equal-time component of the retarded interactions in Eq.(\ref{eq:density}) contribute constants  to the system Hamiltonian $[(n_i-\frac 12)^2=\frac 14]$; thus the bath does not renormalize the Hamiltonian parameters in the system.  Experimentally, in the hybrid nanowires, the Ohmic dissipation can be realized via an electrostatic coupling of quantum wire to metallic gates/films\cite{Cazalilla2006}, while in the ultracold atomic setup, a three dimensional Fermi sea can be considered as a microscopic realization of such an Ohmic environment\cite{Malatsetxebarria2013}.

 Owing to the retarded interactions in Eq.(\ref{eq:density}), our model is a genuine interacting system even though the system Hamiltonian.(\ref{eq:Kitaev}) itself is quadratic. The first step we take to solve this model is performing the Jordan-Wigner transformation(JWT) to map the Kitaev model into a transverse Ising (TI) model:
$H_s=-J\sum_i \sigma^x_i\sigma^x_{i+1}-\frac{\mu}2\sum_i \sigma^z_i$  ($\sigma_i$ the Pauli matrices).
This enables us to study this model via QMC simulations with the worm algorithm~\cite{Prokofev1998} even in the presence of retarded interaction(see the Suppl. Mat.~\cite{Supplementary}), which is invariant under JWT (with  $n_i-\frac 12$ replaced by $\frac 12\sigma_i^z$). Compared to the standard worm QMC algorithm~\cite{Pollet2005}, the only difference here is the calculation of the integrals resulting from the retardation and including them into the QMC acceptance ratio during the updates of the samplings\cite{Cai2014}. For some special quantum models with exact higher dimensional classical mapping, the dissipation can be studied by classical QMC simulations~\cite{Werner2004,Werner2005a,Werner2005b,Sperstad2011,Stiansen2012}.    Recently, lattice systems with retarded interaction in imaginary time have also been studied by determinant QMC\cite{Assaad2007,Hohenadler2012} and directed-Loop QMC algorithm\cite{Weber2017}. We focus on the ground state $(T = 0)$ of the total system. In our QMC simulations, the inverse temperature is scaled as $\beta=L$, corresponding to a dynamical critical exponent $z=1$, which is indeed the case in the QPT in the dissipationless TI model. The periodic boundary condition (PBC) in our simulations corresponds to PBC/anti-PBC in the Kitaev wire depending on the odd/even parity of the particle number. Our model preserves the parity of the particle number of fermions even in the presence of dissipation, which allows us to restrict our measurement in the even parity subspace, which corresponds to ground state of finite system.

\begin{figure*}[htb]
\includegraphics[width=0.95\linewidth]{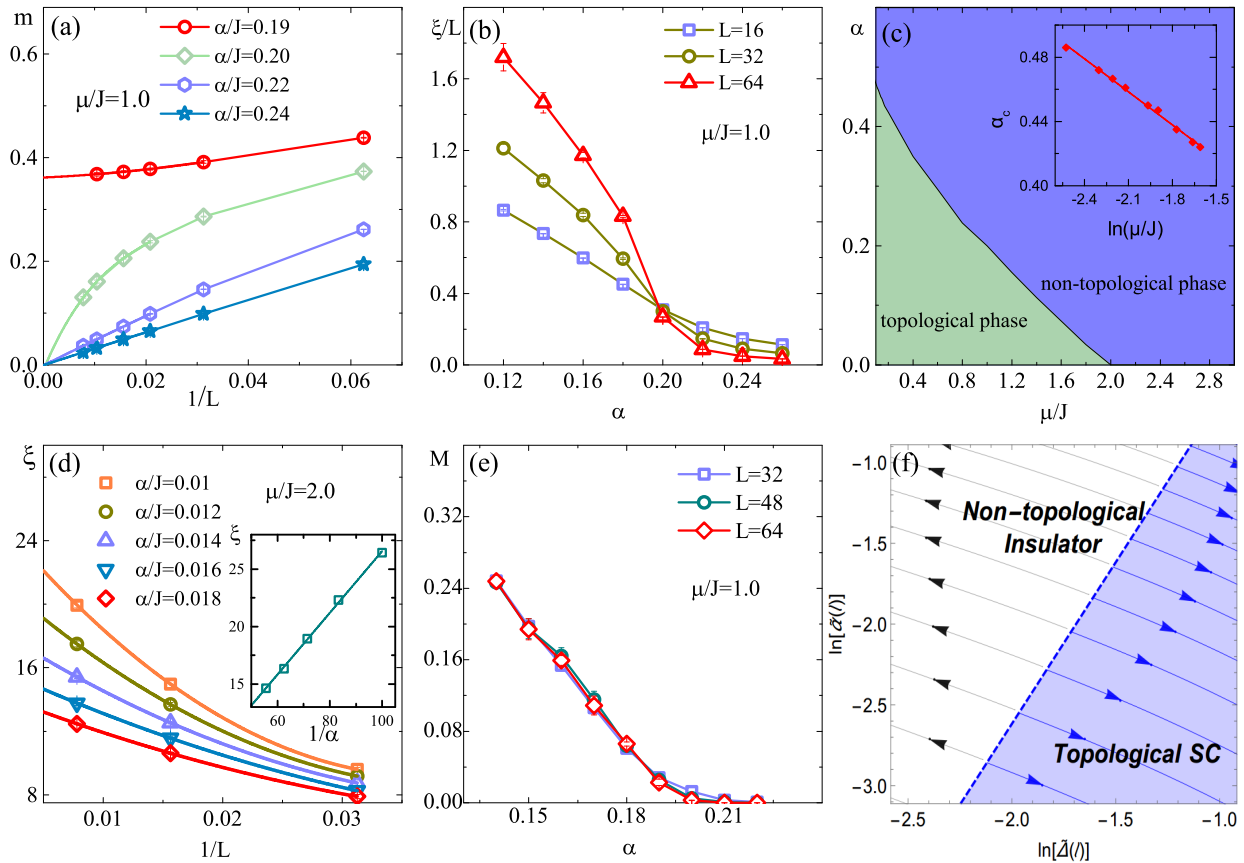}
\caption{(a) Finite size scaling of the structure factor with different $\alpha$; (b) correlation length normalized by the size $L$ as a function of $\alpha$;  (c) phase diagram of the dissipative Kitaev model (or the equivalent dissipative TI model), the inset shows that for small $\mu$ the phase boundary satisfies the relations $\alpha_c\sim \ln\mu$, as predicted by the perturbation theory; (d)finite size scaling of the correlation length with different $\alpha$ values near the critical point $\mu_c=2J$ of the dissipationless TI model; (the inset shows the correlation length as a function of $1/\alpha$ at $\mu_c=2J$) (e) dissipation($\alpha$) dependence of the correlation function between the Majorana fermions at the two ends of the chain; (f) RG flow diagram for $\tilde{\Delta}(\ell)$ and $\tilde{\alpha}(\ell)$ with an initial $K_0=0.501$, the dashed blue line satisfies $\frac{dK(\ell)}{d\ell}=0$ (e.g. condition $2\pi\tilde{\Delta}^2=K^2\tilde{\alpha}$). [$\mu=J$ for (a),(b) and (e)] and $\beta=L$ .}
\label{fig1}
\end{figure*}

{\it Phase diagram and dissipation-driven quantum criticality:}
For the dissipationless case($\alpha=0$), it is well known that the ground state of Hamiltonian.(\ref{eq:Kitaev}) experiences a QPT from a topological nontrivial phase to a trivial one at $\mu=2J$.  In the following, we will focus on the topological non-trival phase (e.g. $\mu=J$) and investigate its fate with increasing dissipation.  Since the QMC with worm update only apply for bosonic or spin systems, what we actually simulate is a TI model with retarded interaction and use its phase diagram to interpret that of the dissipative Kitaev model.  Since both the JWT and Gaussian integral are exact, these two models are exactly equivalent, and thus share the same phase diagram.

We first fix the value of $\mu=J$ and increase $\alpha$. Under the JWT, the topological phase in the Kitaev model can be mapped onto a magnetically ordered phase with spontaneous $Z_2$ symmetry breaking; therefore, we use the long-range correlation functions $\langle \sigma^x_i\sigma^x_j\rangle$ and their Fourier components $S(Q)=\frac {1}{L^2} \sum_{ij} e^{iQ(i-j)}\langle\sigma^x_i\sigma^x_j \rangle$ (structure factor) to identify the QPT induced by dissipation. We define $m=\sqrt{S(Q=0)}$ as the order parameter of the magnetic ordering phase, which extrapolates to its ground state value $m_0$ as $L=\beta\rightarrow \infty$ in  finite size scaling.  As shown in Fig.\ref{fig1}(a), for small $\alpha$, $m_0$ is finite, while it vanishes in the presence of large dissipation. This dissipation-driven QPT can be further verified by the correlation length $\xi$, which can be calculated from the structure factors $S(Q)$ at $Q=0$ and $Q=2\pi/L$\cite{Sandvik2010}. The normalized correlation length $\xi/L$ as a function of $\alpha$ for different system sizes has been plotted in Fig.\ref{fig1} (b), where we can find a crossing point, indicating a scaling invariant quantum critical point (QCP). As shown in Fig.\ref{fig1} (c), there are two distinct phases in the phase diagram of this model: a ferromagnetic phase (or topological phase in the fermonic language) and a paramagnetic phase.

More details about the dissipation-driven QPT can be found by comparing the role of dissipation with that of temperature(T), since both of them tend to suppress quantum fluctuations. It is well known that the TI model is a prototype model to illustrate quantum critical matter, whose properties are determined by the QCPs even at a finite temperature\cite{Hertz1976,Millis1993,Coleman2005}. The question is what happens if the finite T is replaced by dissipation?  (near the QPC, we increase dissipation but fix the temperature of the total system to be zero).  To study this problem, we focus on the QCP of the dissipationless TI model at $\mu=2J$, and calculate the dependence of the spatial correlation length ($\xi$) on $\alpha$ in the case of weak dissipation.  As shown in the inset of Fig.\ref{fig1} (d), $\xi$ is proportional to $1/\alpha$ for weak dissipation, similar as the temperature dependence of $\xi$ in quantum critical regime at finite T\cite{Sachdev1999}. Therefore, the dissipation plays a similar role as temperature near the QCP, while a qualitative difference is that in 1D, the long-range magnetic order is fragile at any finite T, but robust against small dissipation.

{\it Crossover at $\mu=0$: a symmetry protected topological phase;} In the $\alpha-\mu$ phase diagram, the line $\mu=0$ is special as the total Hamiltonian $H_{tot}$ with $\mu=0$ possesses extra symmetries besides the parity symmetry (e.g. $[\mathcal{\hat{S}},H_{tot}]=0$ with   $\mathcal{\hat{S}}=\prod_i\sigma^z_i$). At $\mu=0$, at each site $i$,  $H_{tot}$ is invariant under a combined transformation defined as $\hat{\mathcal{P}}_i=\hat{\sigma}_i^x\otimes_k \hat{P}_{ik}$, where $\hat{P}_{ik}$ is the inversion operator for the $k$th mode harmonic oscillator at site i: $\hat{P}^{-1}_{ik} X_{ik}\hat{P}_{ik}=-X_{ik}$. It is easy to check that each $\mathcal{\hat{P}}_i$ commutes with $H_{tot}$ ($[\mathcal{\hat{P}}_i,H_{tot}]=0$), indicating infinite number of conserved quantities. Even though both $\mathcal{\hat{P}}_i$ and $\mathcal{\hat{S}}$ commute with $H_{tot}$, they don't commute with each other $[\mathcal{\hat{S}},\mathcal{\hat{P}}_i]\neq 0$, which indicates that all the eigenstates are at least doubly degenerate. In Josephson junction arrays\cite{Loffe2002,Doucot2005} and trapped ions\cite{Milman2007}, similar degenerate states with noncommutative conserved quantities have been proposed to be used to construct topologically stable qubits that are robust against decoherence. In our model, these extra symmetries and degeneracies at $\mu=0$ will give rise to remarkable consequences, as we will show in the following.

We focus on the strongly dissipative limit, and perform a perturbation analysis of the total Hamiltonian $H_{tot}$. In the case of $\mu=J=0$, different lattice sites are decoupled and for each site, the groundstate are doubly degenerate, denoted as ``dressed'' spin states ($|\tilde{\uparrow}\rangle_i$ and $|\tilde{\downarrow}\rangle_i$) satisfying the relation $|\tilde{\uparrow}\rangle_i=\hat{\mathcal{P}}_i |\tilde{\downarrow}\rangle_i$. In the strong dissipative limit $\{J,\mu\}\ll \{\lambda_k,\omega_k\}$, one can consider the ``system'' Hamiltonian $H_s$ as perturbations, and derive an effective Hamiltonian $\tilde{H}$ in the $2^L$-dimensional constraint Hilbert spaces spanned by the $\{\tilde{\sigma}^z_i\}$ eigenbasis of the ``dressed'' spin (see the Suppl.Mat for details). In the 1st order perturbation, the effective Hamiltonian can be written in terms of the Pauli operators of the ``dressed'' spin $\tilde{\mathbf{\sigma}}_i$ as:  $\tilde{H}=\sum_i [-\tilde{J}\tilde{\sigma}^x_i\tilde{\sigma}^x_{i+1}-\frac{\tilde{\mu}}2\tilde{\sigma}_i^z]$, where the effective coupling is strongly suppressed by dissipation $\tilde{J}=(\frac a\Omega)^\alpha J$  with $\Omega$ and $a$ the ultraviolet and infrared frequency cutoff of the bath (see Suppl. Mat.~\cite{Supplementary}), while the chemical potential is not $\tilde{\mu}=\mu$. This perturbative results indicates that at strongly dissipative limit, the phase boundary occurs at $\alpha_c\sim -\ln\mu$, which agrees with our numerical results. Another prediction is the absence of quantum phase transition at $\mu=0$, indicating that at this point, dissipation can not completely destroy the topological phase at zero temperature. This robustness is related with the special symmetries and  infinite conserved quantities even in the presence of dissipation, as we analyzed above.

{\it Fate of Majorana edge mode in the presence of dissipation:} Up to now, our discussion was based on spin models.  Even though the long-range magnetic correlations can be considered as an indicator of the topological phase in the fermionic counterpart under JWT; they are not directly physically observable in Kitaev model since they involve nonlocal correlations of string operators in terms of fermion operators. In general, a topological phase is characterized by distinct integer values of topological invariant quantities. However, for an interacting open quantum system as in our case, it is challenging to define or calculate such a topological invariant quantity. An alternative feature of a topological phase is the existence of robust zero modes localized at the edges, known as Majorana edge mode in the Kitaev model. The existence of Majorana mode is characterized by the nonvanishing correlations between the Majorana fermions defined at two ends of the 1D lattice with open boundary condition: $\mathcal{M}=-i\langle \gamma_1\gamma_{2L}\rangle$ with $\gamma_{2i-1}=c_i+c^\dag_i$ and $\gamma_{2i}=i(c^\dag_i-c_i)$ the Majorana fermion operators.

Understanding the effect of the environment on Majorana fermions is crucial and of practical significance for current experiments in solid-state devices\cite{Mourik2012,Deng2012,Churchill2013,Nadj2014,Sun2016,He2017,Zhang2018}. A variety of theoretical methods have been employed to study this problem under various approximations\cite{Goldstein2011,Rainis2012,Pedrocchi2015,Hu2016,Knapp2016,Liu2017}, most of which focus on the dynamical aspect of the environment, modeled by classical noise that heat the system and destroys the topology via a crossover.   Here, we focus on the other aspect, dissipation, of environment, which is relevant for the low-temperature steady state properties.  Recently, the effect of dissipation on the tunneling of Majorana fermion has been discussed analytically\cite{Matthews2014}. Here, We calculate the quantity $\langle \gamma_1\gamma_{2L}\rangle$ and use it to characterize the topological phase and Majorana edge mode in our dissipative Kitaev model. In our QMC simulations this quantity can be expressed in terms of the spin operators  $\mathcal{M}=-i\langle \gamma_1\gamma_{2L}\rangle=\langle \sigma_1^x\sigma_L^x \prod_{i=1}^L\sigma_i^z\rangle$. The QMC measurement is restricted to the even parity subspace.  $\mathcal{M}$ as a function of $\alpha$ for different system sizes is shown in Fig.\ref{fig1} (e), which reveals that $\mathcal{M}$ vanishes at a critical $\alpha_c$, whose value agrees with the QCP identified by the correlation lengths. In summary, the fate of Majorana edge modes in the presence of dissipation indicates that it will drive a topological nontrivial phase into a trivial one without Majorana edge mode via a continuous QPT.

{\it Bosonization analysis:} To get a better understanding of the dissipation-driven QPT, we first consider the continuum limit and use the bosonization technique  to analyze the effective field theory of our model. Following the standard bosonization procedures\cite{Giamarchi2003}, the spinless fermion operator can be decomposed as $\psi(x)=e^{-ik_Fx}\psi_L(x)+e^{ik_Fx}\psi_R(x)$, where the right(left)-moving operator $\psi_{L(R)}(x)$ can be expressed in terms  bosonic operators $\phi(x)$ and $\theta(x)$ as: $\psi_{R/L}(x)=\frac{U_r}{\sqrt{2\pi a}}e^{\mp i \phi(x)+i\theta(x)}$ where $U_r$ is the Klein factor and $a$ is the short distance cutoff. The density operator of the fermions can be expressed as $\rho(x)=-\frac 1\pi\nabla \phi(x)+\frac 1{2\pi\alpha} [e^{2i(k_F x-\phi(x))}+h.c.]$ with  $\phi(x)$ and $\theta(x)$ satisfying the relation $[\phi(x),\nabla\theta(x')]=i\pi \delta(x-x')$.

In the continuum limit, the system Hamiltonian can be considered as a 1D superconductor with p-wave pairing, which can be expressed in terms of $\phi(x)$ and $\theta(x)$:\cite{Lobos2012b}.
\begin{equation}
H_s=\int dx \big\{ \frac{v}{2\pi}[\frac{1}{K}(\partial_x \phi)^2+K(\partial_x\theta)^2]+\tilde{\Delta}\rho_0\sin(2\theta) \big\} \label{eq:pwave}
\end{equation}
where $K$ is the Luttinger parameter, $v$ is the sound velocity, and  $\tilde{\Delta}$ is a dimensionless parameter characterizing the strength of the p-wave pairing.  We focus on the incommensurate filling case which allows us to ignore the spatially fast oscillating terms. The effective action describing the retarded interaction induced by dissipation can also be expressed in the bosonization language\cite{Cazalilla2006}:
\begin{equation}
S_{\textrm{ret}}=-\frac{\tilde{\alpha}}{a_0} \int dx \int d\tau d\tau' \frac{\cos
2[\phi(x,\tau)-\phi(x,\tau')]}{(\tau-\tau')^2}.
\end{equation}
where the dimensionless parameter $\tilde{\alpha}$ is proportional to the dissipation strength. Again, we ignore the higher order irrelevant terms such as $\int dq d\omega |\omega|q^2
|\phi (q,\omega)|^2$. As a consequence, the effective action of the dissipative Kitaev model can be written as below.
\begin{equation}
S_{\textrm{eff}}=\int_0^\beta d\tau[\int dx \frac{1}{i\pi}\dot{\theta}\partial_x\phi+H_s(\tau) ]+S_{\textrm{ret}}
\end{equation}

To study the interplay between the dissipation and p-wave pairing, we perform the standard perturbative renormalization group(RG) procedure to analyze the RG-flow of the parameters $\tilde{\alpha}$, $\tilde{\Delta}$, $v$ and $K$  , and their flow equations read (see the Suppl. Mat.~\cite{Supplementary} for detailed derivations):
\begin{eqnarray}
\nonumber \frac{d\tilde{\Delta}(\ell)}{d\ell}&=&[2-\frac 1{K(\ell)}]\tilde{\Delta}(\ell),\\
\nonumber \frac{d\tilde{\alpha}(\ell)}{d\ell}&=&[1-2K(\ell)]\tilde{\alpha}(\ell)\\
\nonumber \frac{dv\left(\ell\right)}{d\ell}&=&-2\pi K\left(\ell\right)v\left(\ell\right)\tilde{\alpha}\left(\ell\right),\label{eq:RG_v}\\
\frac{dK\left(\ell\right)}{d\ell} &=&4\pi^2\tilde{\Delta}^2\left(\ell\right)-2\pi K^{2}\left(\ell\right)\tilde{\alpha}\left(\ell\right). \label{eq:RGflow}
\end{eqnarray}
 The main results of our model can be illustrated by the flow equations  Eq.(\ref{eq:RGflow}), from which we can find a phase transition point at $K_c=1/2$. For $K(\ell)>K_c$, $\tilde{\Delta}(\ell)$ flows to the strong coupling limit while $\tilde{\alpha}(\ell)$ goes to zero.  $\tilde{\Delta}(\ell)$ couples to the $\sin2\theta$ terms in the Hamiltonian(\ref{eq:pwave}); once it become relevant the field $\theta$ becomes pinned to one of the two degenerate energy minima of the potential:  $\theta=-\pi/4$ or $3\pi/4$, indicating a spontaneous $Z_2$ symmetry breaking observed in our QMC simulations for small $\alpha$. For $K(\ell)<K_c$, the dissipation is relevant while the effect of the pairing is suppressed in the RG sense, therefore this phase can be understood as a dissipative Luttinger liquid which has been investigated analytically~\cite{Neto1997,Cazalilla2006,Malatsetxebarria2013} and numerically\cite{Cai2014}. The intertwined effects between the dissipation and pairing can be found from the RG flow equations of $K$ and $v$, from which we can find that the velocity is only renormalized by dissipation, since it essentially breaks the Lorentz invariance of the Luttinger Liquid term.  Dissipation makes the plasmon velocity becomes slower, similar effect has been discussed in the Coulomb drag~\cite{Cazalilla2006,Lobos2011}. By solving the RG flow equations, we plot the RG flow diagram as shown in Fig.\ref{fig1} (f), which shows the diverging RG flows of the dissipation  and pairing parameters in the different regions of the phase space.

{\it Experimental realizations and parameters:} Our results may be relevant with current experiments of topological superfluid and Majorana fermions in both solid state and ultracold atomic setups. In most cases of solid state experiments, the strength of dissipation is difficult to be controlled and tuned,  so as an experimental realization of the dissipative Kitaev model, we follow the implementation of a topological superfluid proposed by Nascimbene~\cite{Nascimbene2013}, and estimate the relevant parameters in corresponding ultracold atomic setups. As proposed in Ref.~\cite{Nascimbene2013}, the 1D Kitaev model can be realized by loading 1D gas of fermionic atoms (e.g.$^{161}$Dy) into a spin-dependent optical superlattice immersed in an environment composed of a two-dimensional(2D) condensate of Feshbach molecules. In an optical lattice with wavelength $\lambda=530nm$ and the lattice depth along x-direction $V_x=5E_r$ ($E_r=\frac{h^2}{2m\lambda^2}=4.38(2\pi)$~kHz is the recoil energy in this setup), by tuning the scattering length of the fermions and the density of the 2D molecules, one can realize the Kitaev model with parameters $J\sim \Delta\approx 0.1E_r$, which corresponds to an energy of $k_B\times21$nK. The 2D condensate of Feshbach molecules induces the attractive interactions between the 1D fermions,  the static or momentum-dependent part of this bath-induced interaction  gives rise to an p-wave pairing, while the dynamical or frequency-dependent part plays a role of quantum dissipation. It has been shown that the quantum environment composed of Bogoliubov quasiparticles in a 2D condensate can give rise to Ohmic dissipation\cite{Torre2010}. By tuning the scattering length between the fermions and molecules, one can realize a dissipation strength $\alpha$ comparable with $\Delta$ and $J$. One of the major  challenges in the experimental implementation  is the finite temperature effect: to observe the dissipation-induced phase transition, the temperature needs to be lower than $20$nK, still below the current experimental limit of the cold fermionic systems.

{\it Conclusion and outlook:} In summary, we have studied the effect of dissipation on topological quantum phases by considering a specific model of Kitaev quantum wire with onsite Ohmic dissipation and found that the topological phase in this model will eventually be destroyed via either  a continuous QPT or a crossover depends on the symmetry of the system. Some avenues for further investigations can be suggested. The first and most important question is the generality of the above results, whether it applies to other topological models with different kind of dissipation. An important feature of our model is that a system particle interacts with the bath via its density operators, this dissipation process preserves the total number (also the parity) of the particles in the system. We expect that our results hold for this type of symmetry-protected dissipation, while for other dissipation mechanisms ({\it e.g.} the particle loss) that break these symmetries, the conclusion may be different. This point needs to be verified numerically, which requires new methods and models\cite{Yan2018}. Another important ingredient still missing is a proper definition of a topological invariant (an integer number) for these interacting open quantum systems, which may provide more direct evidence of the topological phases and topological QPT compared to the existence of edge modes. This topological number needs not only to be well-defined, but also computable in our practical numerical simulations. Last but not the least, our work also rises an interesting question that whether non-trivial topological properties could only exist in a subsystem of reduced dimensionality spatially embedded in a larger non-topological system with an inhomogeneous Hamiltonian, if so, how to identify this subsystem topological phases and what distinguish them from the conventional topological matters?

{\it Acknowledgements -- }  ZC acknowledges the support from the National Key Research and Development Program of China (grant No.2016YFA0302001), the National Natural Science Foundation of China under Grant No.11674221, and the Shanghai Rising-Star program. This work is also supported by the Program for Professor of Special Appointment (Eastern Scholar) at Shanghai Institutions of Higher Learning. A.M.L. acknowledges financial support from PICT-0217 2015 and PICT-2017 2018 (ANPCyT - Argentina), PIP-11220150100364 (CONICET - Argentina) and Relocation Grant  RD1158 - 52368 (CONICET - Argentina). We acknowledge the support from the Center for High Performance Computing of Shanghai Jiao Tong University.


\end{document}